\PassOptionsToPackage{pdftex}{graphicx}
\documentclass[shortnote,twocolumn]{jpsj3}
\usepackage{graphicx,bm,amssymb,amsmath,url}
\usepackage{tikz}
\usepackage[colorlinks=true,allcolors=blue]{hyperref}

\newcommand{\pb}{\bar p}

\makeatletter
\let\balancecolumns\relax
\let\balancing@outputdblcol\relax
\let\balancing@outputpage\relax
\providecommand\ext@figure{lof}
\providecommand\ext@table{lot}
\def\@lastpagebalancing{false} 
\makeatother
\title{Graph Partitioning Based on Pair Interaction Energy Using Quantum Annealer}

\author{
    Masayuki \textsc{Ohzeki}\thanks{E-mail: mohzeki@tohoku.ac.jp}
}

\inst{
    Graduate School of Information Sciences, Tohoku University, Sendai 980-8579, Japan\\
    Department of Physics, Institute of Science Tokyo, Meguro, Tokyo 152-8551, Japan\\
    Research and Education Institute for Semiconductors and Informatics, Kumamoto University, Kumamoto 860-8555, Japan\\
    Sigma-i Co., Ltd., Minato, Tokyo 108-0075, Japan
}

\title{A Minimal Duality Estimate for the Surface-Code Threshold under Nearest-Neighbor Correlated Errors}

\abst{
We apply the single-equation duality criterion to the square-octagonal random-bond Ising model recently obtained from an exact error-edge map for a surface code with nearest-neighbor correlated errors.  The calculation is performed for the minimal cell after the error-edge reduction.  
For the symmetric case \(p_1=p_2=p_3=p\), this gives \(p_c=0.0288427147\), in close agreement with the reported numerical threshold of about \(3\%\).
}

\kword{quantum error correcting code, duality, critical point}

\begin{document}
\makeatletter
\let\balancecolumns\relax
\let\balancing@outputdblcol\@outputdblcol
\makeatother
\maketitle

A key challenge in random spin systems is determining critical points.  Among various analytical tools, duality analysis stands out as an unexpectedly powerful and conceptually simple method.  A classic example is the Kramers--Wannier duality for the two-dimensional Ising model~\cite{Kramers1941}, which reveals the exact location of the critical point without requiring the explicit solution of the model.  Duality has since been extended to \(Z_q\) models via Fourier transformation~\cite{WuWang1976}, spin glasses through the replica method~\cite{Nishimori1979,Nishimori2002,Maillard2003,Ohzeki2009regular,Ohzeki2015}, and the theoretical limits of quantum error-correcting codes~\cite{Dennis2002,Ohzeki2009color,Ohzeki2012,Ohzeki2012duality, Bombin2012}.   
The same Fourier-duality viewpoint has also recently been extended to symmetric-group models and random tensor networks~\cite{Ohzeki2024}.

Recently, they introduced an error-edge map (EEM) for the surface code under independent single-qubit \(Z\) errors and nearest-neighbor correlated \(ZZ\) errors~\cite{Wang2026}.  Let \(E\) be the actual error and \(E'\) be a recovery operator with the same syndrome.  Error correction succeeds when \(E+E'\) is homologously trivial, and fails when \(E+E'\) contains a nontrivial logical string.  For a fixed syndrome \(s\), the maximum-likelihood decoder compares the total probabilities \(P_\gamma(s)\) of the possible homology sectors \(\gamma\).  This is the same logic as the standard mapping of the surface-code decoding problem to a random-bond Ising model (RBIM)~\cite{Dennis2002}: a reference recovery fixes the syndrome, and any other recovery with the same syndrome is generated by adding closed loops, represented by Ising spin flips on the dual lattice.

In that mapping, an error variable on an edge becomes a random sign of an Ising coupling.  If \(\ell_e=0,1\) denotes the absence or presence of an effective error on edge \(e\), then
\begin{equation}
\tau_e=(-1)^{\ell_e}
\end{equation}
is the quenched bond variable.  A spin configuration \(\{\sigma\}\) specifies which closed loop is added to the reference recovery; for an edge \(e\) connecting two Ising spins \(a(e)\) and \(b(e)\), the relative loop variable is \(\sigma_{a(e)}\sigma_{b(e)}\).  Thus an error edge contributes a factor
\begin{equation}
\exp\left[K_e \tau_e \sigma_{a(e)}\sigma_{b(e)}\right],
\end{equation}
which is precisely the Boltzmann factor of an Ising bond with a random sign, where \(K_e=K_i\) for \(e\in E_i\): \(\tau_e=+1\) is ferromagnetic and \(\tau_e=-1\) is antiferromagnetic.  This is the sense in which the decoding problem is an RBIM on the graph defined by the error edges.  The EEM of Ref.~\cite{Wang2026} performs this construction exactly for the nearest-neighbor correlated-error model by replacing the original correlated physical processes with effective error edges on the square-octagonal lattice.  Compared with the conventional independent-error surface-code mapping~\cite{Dennis2002}, the resulting RBIM has three edge types and a local EEM constraint.

For a homology sector \(\gamma\), we therefore write the square-octagonal RBIM partition function as
\begin{equation}
Z_\gamma[\tau]=
\sum_{\{\sigma\}}^{\prime}
\exp\left[
\sum_{i=1}^3\sum_{e\in E_i}
K_i \tau_e \eta_{\gamma,e}
\sigma_{a(e)}\sigma_{b(e)}
\right],
\label{eq:statmodel}
\end{equation}
where \(E_i\) denotes the set of effective square-octagonal edges of type \(i\).  The factor \(\eta_{\gamma,e}=\pm1\) specifies the homology sector: \(\eta_{\gamma,e}=1\) for the trivial sector, while changing the sign along a noncontractible cut inserts the domain wall associated with \(\gamma\).  The prime indicates the local EEM constraint \(\sigma_{r}^{1}\sigma_{r}^{2}\sigma_{r}^{3}\sigma_{r}^{4}=1\) in each elementary square-octagonal cell.  The disorder distribution of this RBIM is
\begin{equation}
P_i(\tau_e)=(1-\pb_i)\delta_{\tau_e,1}+\pb_i\delta_{\tau_e,-1}
\qquad (e\in E_i),
\label{eq:disorder}
\end{equation}
with type-dependent effective error probabilities \(\pb_i\).

On the Nishimori line, the Boltzmann weight of Eq.~\eqref{eq:statmodel} is proportional to the probability weight of the corresponding error class.  Therefore
\begin{equation}
\frac{P_\gamma(s)}{P_0(s)}
=
\frac{Z_\gamma[\tau]}{Z_0[\tau]},
\qquad
\Delta F_\gamma=-\log\frac{Z_\gamma[\tau]}{Z_0[\tau]},
\label{eq:domainwall}
\end{equation}
up to an overall normalization independent of \(\gamma\).  Here \(0\) denotes the trivial homology sector.  If the domain-wall free-energy cost \(\Delta F_\gamma\) grows with the linear size \(L\), nontrivial logical sectors are suppressed, and the logical error probability under maximum-likelihood decoding vanishes as \(L\to\infty\).  If \(\Delta F_\gamma\) remains finite, a nontrivial logical sector has nonzero weight in the thermodynamic limit and decoding fails with nonzero probability.  The crossing of the phase boundary of Eq.~\eqref{eq:statmodel} with the Nishimori line is therefore precisely the maximum-likelihood error threshold of the quantum code~\cite{Dennis2002,Wang2026}.  In the following local duality calculation, we use the trivial sector, \(\eta_{\gamma,e}=1\), because the principal Boltzmann factor estimates the point at which this sector loses stability against the domain-wall excitation.

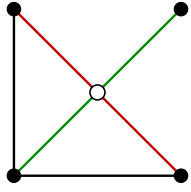
\begin{figure}[t]
\centering
\begin{tikzpicture}[
  scale=0.92,
  fixed/.style={circle,fill=black,draw=black,inner sep=1.8pt},
  summed/.style={circle,fill=white,draw=black,line width=0.6pt,inner sep=2.0pt},
  edge/.style={line width=0.9pt}
]
  \coordinate (A) at (0,0);
  \coordinate (B) at (2.4,0);
  \coordinate (C) at (2.4,2.4);
  \coordinate (D) at (0,2.4);
  \coordinate (O) at (1.2,1.2);

  \draw[edge] (A) -- (B);
  \draw[edge] (A) -- (D);
  \draw[edge,red!75!black] (D) -- (B);
  \draw[edge,green!55!black] (A) -- (C);

  \node[fixed] at (A) {};
  \node[fixed] at (B) {};
  \node[fixed] at (C) {};
  \node[fixed] at (D) {};
  \node[summed] at (O) {};
\end{tikzpicture}
\caption{
(Color online) Minimal cluster entering the principal Boltzmann factor \(x_0\).
Black circles denote endpoints fixed to the trivial sector in the principal-factor estimate, and the open circle denotes the local EEM degree of freedom summed or contracted in the reduced effective-edge description.
Black, red, and green lines represent \(l_1\), \(l_2\), and \(l_3\) effective edges with probabilities \(\bar p_1\), \(\bar p_2\), and \(\bar p_3\), respectively.
The cluster contains two \(l_1\) edges and one each of \(l_2\) and \(l_3\), leading to \(m_1=2\) and \(m_2=m_3=1\).
}
\label{fig:cluster}
\end{figure}

Their mapping gives three classes of effective error edges.  The effective probabilities, denoted here by \(\pb_i\), are
\begin{equation}
\pb_1=p_1,\qquad
\pb_2=2p_2(1-p_2),\qquad
\pb_3=2p_3(1-p_3),
\label{eq:pbar}
\end{equation}
as in Eq.~(20) of Ref.~\cite{Wang2026}.  Here \(p_1\) is the probability of a single-qubit error process, while \(p_2\) and \(p_3\) are the probabilities of diagonal and anti-diagonal correlated error processes.  The EEM cell contains two \(l_1\) edges and one each of \(l_2\) and \(l_3\).  We therefore set
\begin{equation}
m_1=2,\qquad m_2=m_3=1.
\label{eq:multiplicity}
\end{equation}
The minimal cluster used in this counting is shown in Fig.~\ref{fig:cluster}.

The Nishimori condition for each effective edge reads
\begin{equation}
e^{-2K_i}=\frac{\pb_i}{1-\pb_i},
\qquad
K_i=\frac12\log\frac{1-\pb_i}{\pb_i}.
\label{eq:nishimori}
\end{equation}
For \(n\) replicas, the principal Boltzmann factor of an effective edge of type \(i\), corresponding to the trivial relative spin configuration in all replicas, is
\begin{equation}
x_0^{(i)}
=
(1-\pb_i)e^{nK_i}+\pb_i e^{-nK_i}.
\label{eq:x0edge}
\end{equation}
Thus, the principal factor of the minimal EEM cell is
\begin{equation}
x_0=\prod_{i=1}^3\left[x_0^{(i)}\right]^{m_i}.
\label{eq:x0cell}
\end{equation}
The corresponding dual principal factor is obtained by the \(Z_2\) Fourier transform of the cell weight.  Since the reduced EEM cell has one independent \(Z_2\) syndrome variable per replica, the normalization factor is \(2^{-n}\), and
\begin{equation}
x_0^\ast
=
2^{-n}
\prod_{i=1}^3
\left(e^{K_i}+e^{-K_i}\right)^{n m_i}.
\label{eq:x0dual}
\end{equation}
This is the quantity that must be used for the present correlated-error problem.  Treating the two terms generated by an \(l_2\) or \(l_3\) process in the square-octagonal Hamiltonian as independent bonds would change Eq.~\eqref{eq:x0dual}; this would no longer reproduce the standard surface-code condition \(h(p)=1/2\) when the correlated processes are absent.

The single-equation duality estimate is
\begin{equation}
x_0=x_0^\ast .
\label{eq:duality}
\end{equation}
Taking the replica limit gives the desired scalar equation.  From Eqs.~\eqref{eq:x0cell} and \eqref{eq:x0dual}, the limit of the replica number ($ n\to 0$) yields, in conjunction with Eq.~\eqref{eq:nishimori}, 
\begin{equation}
K_i(1-2\pb_i)-\log(e^{K_i}+e^{-K_i})
=
(1-\pb_i)\log(1-\pb_i)+\pb_i\log\pb_i .
\end{equation}
Therefore Eq.~\eqref{eq:duality} becomes, in bits,
\begin{equation}
\sum_{i=1}^3 m_i h(\pb_i)=1,
\qquad
h(x)=-x\log_2x-(1-x)\log_2(1-x).
\label{eq:general}
\end{equation}
Substituting Eqs.~\eqref{eq:pbar} and \eqref{eq:multiplicity}, we obtain
\begin{equation}
2h(p_1)+h\{2p_2(1-p_2)\}+h\{2p_3(1-p_3)\}=1 .
\label{eq:main}
\end{equation}

For the symmetric case considered in Ref.~\cite{Wang2026}, \(p_1=p_2=p_3=p\), Eq.~\eqref{eq:main} reduces to
\begin{equation}
2h(p)+2h\{2p(1-p)\}=1 .
\label{eq:symmetric}
\end{equation}
Solving this equation gives
\begin{equation}
p_c=0.0288427147\ldots.
\end{equation}
The result is close to the threshold of approximately \(3\%\) reported from Monte Carlo simulations in Ref.~\cite{Wang2026}.  It also reduces to the familiar independent-error estimate \(h(p_c)=1/2\), or \(p_c=0.1100278644\ldots\), when only the two \(l_1\) single-qubit error edges are present.

We emphasize that Eq.~\eqref{eq:main} is the minimal-cell duality estimate.  Systematic cluster improvement should be possible in the spirit of real-space-renormalized duality \cite{Ohzeki2008hl, Nishimori2006} and graph-polynomial methods~\cite{Ohzeki2015}.  However, for the constrained square-octagonal representation in Ref.~\cite{Wang2026}, the cluster must preserve the EEM cell and its dual normalization.  A naive toroidal cluster of the unreduced square-octagonal graph can double-count the constrained \(l_2\) and \(l_3\) contributions, leading to a different model estimate.

\begin{acknowledgment}
We received financial support from the Cross-ministerial Strategic Innovation Promotion Program (SIP) of the Cabinet Office (No. 23836436).
\end{acknowledgment}

\end{document}